\documentclass[useAMS,usenatbib,usedcolumn]{mn2e}
\usepackage{longtable}
\usepackage{psfig}
\usepackage{epsfig}
\usepackage{graphicx}
\usepackage{times}

\def\hd{^{\rm h}}
\def\md{^{\rm m}}
\def\sd{^{\rm s}}

\title[A Chandra View of the Multiple Merger In Abell 2744]
{A Chandra View of the Multiple Merger In Abell 2744}

\author[J. C. Kempner \& L. P. David]
{ Joshua C. Kempner\thanks{jkempner@cfa.harvard.edu} and Laurence P. David\\
Harvard-Smithsonian Center for Astrophysics, 60 Garden St.,
Cambridge, MA 02138\\ }
\date{Submitted 2003 October 7}

\begin{document}

\maketitle
\begin{abstract}
We present a {\it Chandra}\/ observation of the merging cluster of galaxies
Abell 2744.  The cluster shows strong evidence for an ongoing major merger
which we believe to be responsible for the radio halo.  X-ray emission and
temperature maps of the cluster, combined with the spatial and redshift
distribution of the galaxies, indicate a roughly north-south axis for the
merger, with a significant velocity component along the line of sight.  The
merger is occurring at a very large velocity, with $\mathcal{M}$~=~2--3.
In addition, there is a small merging subcluster toward the northwest,
unrelated to the major merger, which shows evidence of a bow shock.  A
hydrodynamical analysis of the subcluster indicates a merger velocity
corresponding to a Mach number of $\sim$1.2, consistent with a simple
infall model.  This infalling subcluster may also be re-exciting electrons
in the radio halo.  Its small Mach number lends support to turbulent
reacceleration models for radio halo formation.
\end{abstract}


\section{Introduction}
\label{sec:intro}

One of the largest contributions that {\it Chandra}\/ has made in its first
few years of operation is in the study of merging clusters of galaxies.
The unprecedented spatial resolution of the satellite has made possible the
study of detailed physics of cluster interactions.  Cold fronts---the sharp
leading edges of moving cool cores of gas from clusters---along with their
associated bow shocks have been imaged for the first time using {\it
Chandra}.  From these measurements, the dynamics of cluster mergers have
been determined \citetext{e.g. A2142, \citealp{mpn+00}; A3667,
\citealp*{vmm01b}; 1E0657-56, \citealp{mgd+02}}.  These anlyses at high
spatial resolution have also made it possible to demonstrate the
suppression of conduction in clusters \citep{ef00b, vmm01a}, to determine
the dark matter distribution on small scales \citep{vm02}.  The resolution
of {\it Chandra} has also provided the basis for the first measurement of a
direct correlation between cluster merger shocks and diffuse radio emission
in clusters \citep{mv01}.

Abell 2744, also known as AC 118, is a rich (Abell richness class 3),
luminous \citep[$L_X {\rm (0.1-2.4 keV)} = 22.05 \times 10^{44}$ erg
sec$^{-1}$;][]{evb+96} cluster at moderate redshift
\citep[$z=0.308$;][]{cn84}.  It hosts one of the most luminous known radio
halos which covers the central 1.8 Mpc of the cluster, as well as a large
radio relic at a distance of about 2 Mpc from the cluster center
\citetext{\citealp*{gtf99}, \citealp{gef+01,gfg+01}}.  Because of the
presence of the radio halo and relic, Abell 2744 has been known to be
undergoing a merger, but the details of the merger have been rather murky.
\citet{abe58} classified the spatial distribution of its galaxies as
``regular,'' but it has no dominant bright galaxiy or galaxies so its
Bautz-Morgan class is III \citep{bm70}.

Observations of the cluster with {\it ROSAT}\/ shed some light on the
merger, showing the presence of a second peak in the X-ray brightness a
little less than 1 Mpc to the northwest of the main peak.  This second peak
is much smaller and presumably much less massive than the main cluster,
although it could have been stripped of much of its gas if it had already
passed through the main cluster.  The radio halo extends in the direction
of this second peak, leading to the impression that the merger of the large
main cluster and this smaller subcluster to the northwest is the cause of
the radio halo, perhaps accelerating electrons via turbulence in its wake.

Our observation of the cluster with the higher resolution made possible by
{\it Chandra}\/ disproves this picture for the formation of the halo.
While a merger does indeed appear to be resonsible for the radio halo, in
fact it appears to be created by a merger between two subclusters with a
small mass ratio, while the very small subcluster to the northwest is only
beginning its descent into the potential of the two much larger subclusters
and has only a small effect on the nonthermal emission.

We assume $H_0 = 50$ km s$^{-1}$ Mpc$^{-1}$ and $q_0=0.5$ throughout, for
which $1\arcsec=5.58$ kpc and $d_L = 1970$ Mpc.  All errors are quoted at
90\% confidence unless otherwise stated.

\section{Observation and Data Reduction}
\label{sec:obs}

The data were taken on September 3, 2001 in a single observation of
$\sim$24,800 seconds using the ACIS-S detector on {\it Chandra}, with
the focus on the S3 chip.  The data were taken in Very Faint (VF) mode, in
order to allow for additional rejection of particle background
events\footnote{http://cxc.harvard.edu/cal/Links/Acis/acis/Cal\_prods/vfbkgrnd/index.html}.
Very Faint mode data retains 5$\times$5 pixel islands for each event,
thereby making it possible to identify charged particle events that would
appear to be valid photon events using the standard 3$\times$3 pixel
islands.  This additional filtering of background events was performed,
then the data were filtered on the standard {\it ASCA}\/ grades 0, 2, 3, 4,
and 6.  Observations with {\it Chandra}\/ are frequently affected by
background flares where the background increases over the quienscent level.
We checked for flares using the S-1 chip, which is quite sensitive to
flares and is far enough away from the focal point of the detector to be
essentially devoid of source photons.  No flares occurred during our
observation.

We used the period D blank sky background files for background correction.
The background files were also screened using the VF mode filtering.  We
checked the quiescent background rate in our data by measuring the event
rate on the S-3 chip in PHA channels 2500-3000, where the sensitivity to
photons is extremely small, and compared it to the rate in the same PHA
channels in the blank sky background files.  The background level in our
data was $\sim$19\% below the nominal quiescent background level in the
blank sky files, which include data from the beginning of period D when the
quiescent level was slightly higher than the average for that period.  We
therefore corrected the background by the ratio of these quiescent rates
when subtracting the background in our subsequent analysis.  Despite this
small correction, the background level in our data is consistent with the
nominal quiescent level in PHA channels 2500-3000 at the date of the
observation, to within the observed
scatter\footnote{http://asc.harvard.edu/contrib/maxim/bg/} of the
nominal level.  Throughout this analysis, we used the calibration products
in CALDB 2.17.

\section{X-ray Properties}
\label{sec:global}

\subsection{X-ray Image}
\label{ssec:xray_image}

\begin{figure}
\centerline{\epsfig{width=8.5cm,file=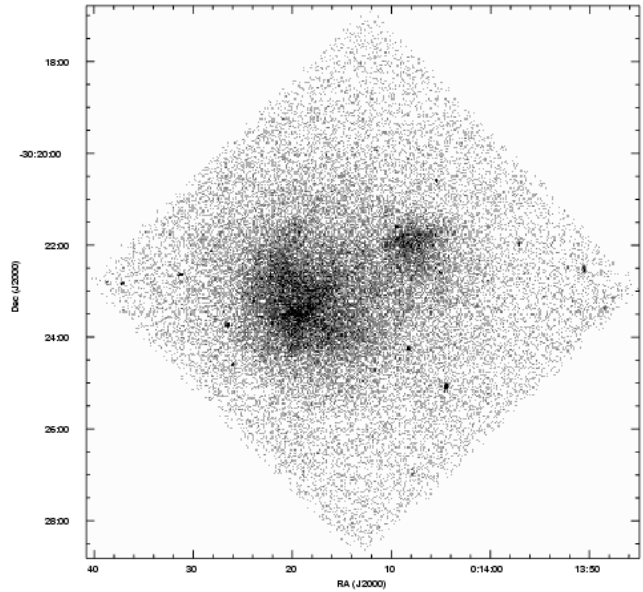}}
\caption{Raw  {\it Chandra} image of Abell 2744.  The entire
S3 chip is shown.
\label{fig:rawimg}}
\end{figure}

\begin{figure*}
\centerline{\epsfig{width=16.0cm,file=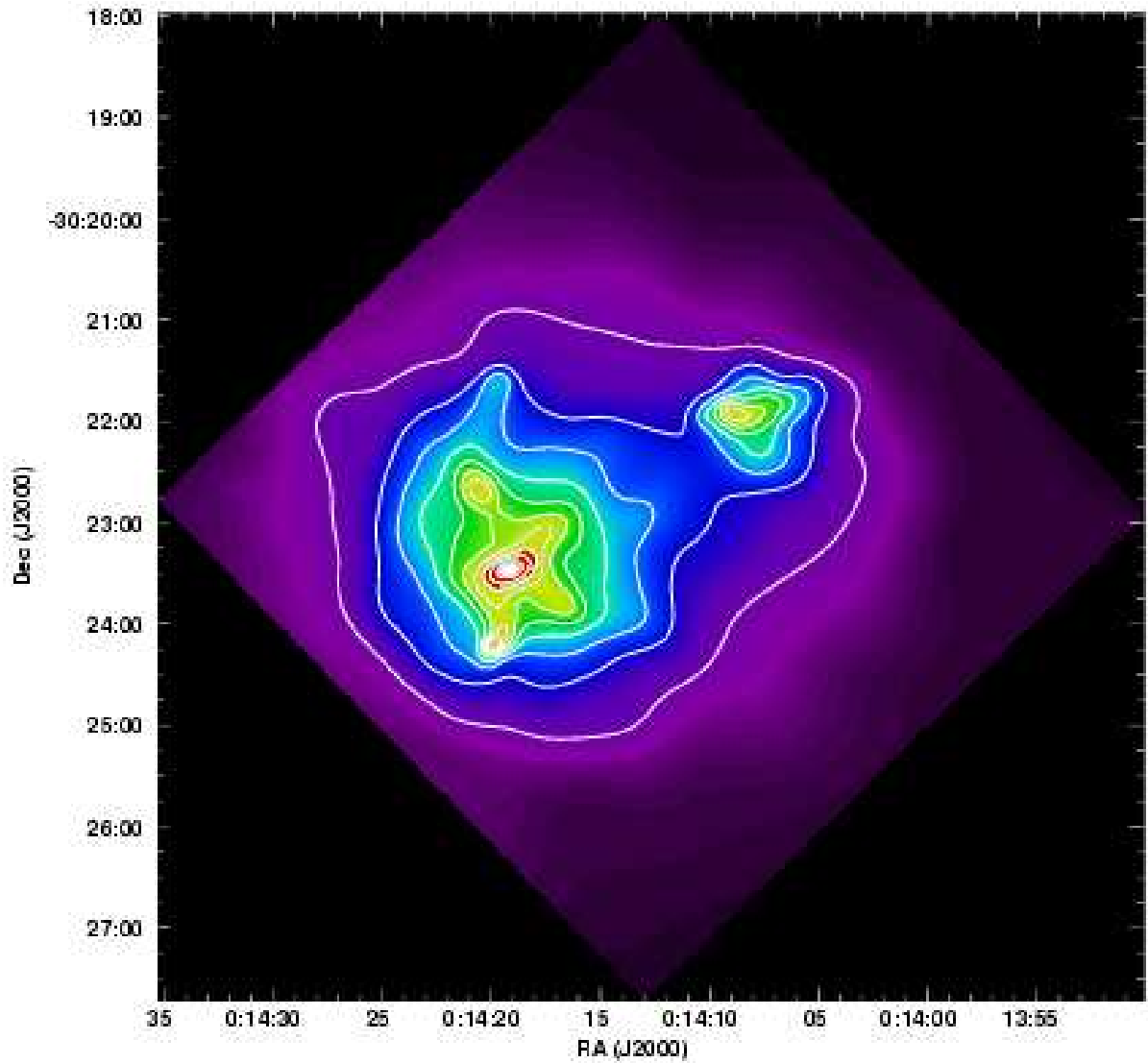}}
\caption{{\it (b)} Adaptively smoothed {\it Chandra} image of Abell 2744.
The point sources were removed before smoothing.  The color scale and contours
are linearly spaced.
\label{fig:smoimg}}
\end{figure*}

We show a raw {\it Chandra}\/ image, binned into 2\arcsec pixels, in
Figure~\ref{fig:rawimg}.  Figure~\ref{fig:smoimg} shows an adaptively smoothed
version of the same image.  The adaptively smoothed image was constructed
using the CIAO tool {\it csmooth} with a minimum signal-to-noise of 3 and a
maximum of 5.  The blank sky background image and exposure map were both
smoothed using the same kernel as for the source image.  These smoothed
images were then used to correct the source image.  A mono-energetic
exposure map was used, with energy at 0.8 keV---the peak of the emission in
the central 2\farcm25 of the cluster.

The {\it Chandra}\/ image shows two distinct components to the cluster: a
large, irregular main cluster, and a much smaller subcluster to its
northwest (see Figure~\ref{fig:smoimg}).  The main cluster, while fairly
strongly peaked, does not fall off in surface brightness uniformly in all
directions.  Rather, it shows four ``ridges'' of emission extending to the
north, northwest, southwest, and south.  These ridges provide strong
evidence, even in the absence of other information, that the cluster is out
of hydrostatic equilibrium.  All the ridges except the south ridge extend
at least an arcminute out of the center, while the north rigde extends
nearly two arcminutes.  These three ridges also show significant curvature,
possibly suggesting internal angular momentum and turbulence within the
main cluster.  Even the ridges are not completely continuous: the south and
north ridges have secondary surface brightness peaks away from the central
peak of the main cluster, at roughly R.A.~=~$00\hd14\md19.5\sd$,
Dec.~=~$-30^\circ24\md09\sd$ and R.A.~=~$00\hd14\md20.5\sd$,
Dec.~=~$-30^\circ22\md45\sd$ respectively.  These secondary peaks are
clearly visible in Figure~\ref{fig:smoimg}.  We suspect these may be
identified with the cool cores of merging subclusters, as we will discuss
below.

At large radii, the substructure in the main cluster lessens.  By a
distance of $\sim$2.1\arcmin\ from the cluster center, the surface
brightness distribution falls off approximately uniformly in all directions
except that of the northwest subcluster.  In a $\beta$-model fit to a {\it
ROSAT}\/ PSPC image of the cluster, \citet{gfg+01} find both an unusually
large core radius of $\sim$640 kpc and an unusually steep value for $\beta$
of $1.0$.  Their fit excludes the subcluster.  The core they find
encompasses exactly the ``ridges'' in the center of the main cluster as is
evident from the {\it Chandra} data.  The core radius they found may also
have been broadened slightly by the large point spread function of {\it
ROSAT}.  We have fit a radial surface brightness profile, centered on the
same point as that used in \citet{gfg+01} (R.A.~=~$00\hd14\md18.7\sd$,
Dec.~=~$ -30^\circ23\md16\sd$), with a single-component $\beta$-model.  We
find the best-fit parameters to be $r_c = (474 \pm 33)$~kpc and $\beta =
0.79 \pm 0.06$ (1$\sigma$ errors).  The {\it Chandra} image shows that the
cluster is too far from equilibrium, particularly in its core, for the
parameters of the $\beta$-model fit to have any useful physical meaning.

About 2.7\arcmin\ ($\sim$900 kpc projected distance) to the northwest of
the main cluster is a much smaller subcluster.  At that distance, the
subcluster is still well within the X-ray halo of the main cluster, which
extends out to a radius of at least 11\arcmin\ as detected by {\it ROSAT}
\citep{gfg+01}.  The subcluster exhibits a sharp, curved eastern edge that
is particularly distinct on its northeast side, with a surface brightness
contrast across the edge of $1.7\pm0.2$ (1$\sigma$) and is hence very
statistically significant..  From there, it fans out to the west, getting
more diffuse with increasing distance from the east edge.  These features
indicate that the subcluster is moving from west to east, while its gas is
swept back by ram pressure from the ICM of the main cluster.  This
characteristic ``fan'' shape of the subcluster is typical of smaller
subclusters being ram pressure stripped by interaction with a larger
cluster \citetext{cf. \citealp{ksr02,mgd+02}}.  As can be seen in
Figure~\ref{fig:rawimg}, the subcluster's fan of emission is brighter to one
side (the north) than to the other.  The contours in Figure~\ref{fig:smoimg}
are also compressed on this side, indicating that the brightness falls off
more steeply that it does on the south side of the subcluster.  This
morphology is similar to that of the south subcluster in Abell 85
\citep*{ksr02}, which is quite similar in terms of both the mass ratio and
the current separation of the constituent clusters.  The ``bullet''
subcluster in 1E0657-56 \citep{mgd+02} also shows this sort of asymmetry.

\subsection{The Northwest Subcluster}
\label{ssec:subcluster}

\begin{figure}
\centerline{\epsfig{width=8.5cm,file=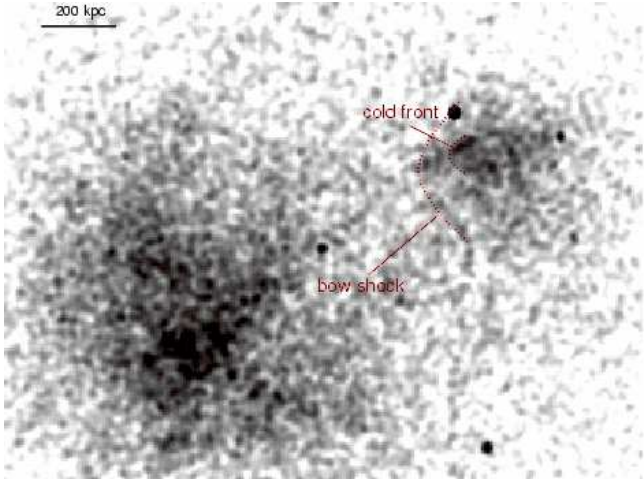}}
\caption{Gaussian smoothed, 0.3-6 keV image of Abell 2744. The cold
front and bow shock are indicated in red.\label{fig:subcluster}}
\end{figure}

As mentioned above, the raw image shows evidence that the northwest
subcluster contains a ``cold front'' similar to those seen by {\it
Chandra}\/ in other clusters \citetext{e.g. A2142, \citealp{mpn+00}; A3667,
\citealp{vmm01b}; 1E0657-56, \citealp{mgd+02}}.  Also visible is a
possible indication of a bow shock about 90 kpc ahead of the cold front
(see Figure~\ref{fig:subcluster}).  Approximating the core as a sphere, the
ratio of the stand-off distance of the shock to the radius of curvature of
the cold core is determined only by the Mach number of the shock.
Following \citet{moe49}, this diagnostic indicates a Mach number of
$\sim$1.2.  The stand-off distance one measures, however, is highly
sensitive to projection.  That is, if the velocity of the subcluster has a
significant line of sight component, we will have underestimated the
stand-off distance, and thereby overestimated the Mach number.  Futhermore,
this method is exact only for a perfect sphere, although the cold front is
unlikely to deviate from this too much, at least on its leading edge.

The photon statistics are too poor to do a properly deprojected
analysis of the electron densities across the cold front.  There are,
however, other methods for measuring the velocity of the subcluster.
The surface brightness jump across the putative bow shock provides an
approximate diagnostic of the subcluster's velocity.  We approximate
the square-root of the ratio of brightnesses on either side of the
shock as the ratio of densities, that is, $(I_{X1}/I_{X2})^{1/2}
\approx \rho_1/\rho_2$, which in turn is equal to the inverse of the
shock compression.  From this we determine the shock compression to be
$C = 1.3\pm0.2$, which implies a Mach number of $\mathcal{M} =
1.2\pm0.2$ (both 1$\sigma$).  From the image we can also roughly
determine the opening angle of the Mach cone to be about 55--60
degrees, which implies a Mach number of 1.15--1.22.  Thus, while the
evidence of a bow shock is certainly not conclusive from the image, the
consistency of the various diagnostics lends credence to the suggestion
of a bow shock.

These various independent diagnostics for the velocity of the
subcluster are all consistent with each other to within the measurement
errors, all finding $\mathcal{M} \approx 1.2$.  This is a good
indication that the merger axis is not highly inclined to the plane of
the sky.  This is quite similar to the velocities found for the other
merging subclusters mentioned above.  As a sanity check, we also
calculated the velocity for a collisionless point mass at the projected
radius of the subcluster, falling from the turnaround radius into an
isothermal potential with a mass of $2.70 \times 10^{15} M_\odot$
\citep{gm01}.  This simplistic model puts the velocity of the test
particle at $\mathcal{M} = 1.2$ given the measured temperature of the
main cluster of $9.3^{+4.9}_{-2.7}$ keV in the vicinity of the
subcluster.  This is also consistent with our measured velocity.  We
should note, however, that this velocity should be treated as an upper
limit for this particular model, since projection effects would
increase the actual separation of the subcluster and decrease its
velocity.  Furthermore, the fan-like shape of the subcluster indicates
that the effects of ram pressure on the gaseous content of the
subcluster are significant, further reducing its likely velocity
compared to our simple model.  Therefore, the model is only useful in
setting the general scale of the subcluster's velocity, and for this
purpose it is completely consistent with our data.

\subsection{Temperature Structure}
\label{ssec:tmap}

\begin{figure}
\centerline{\epsfig{width=8.5cm,file=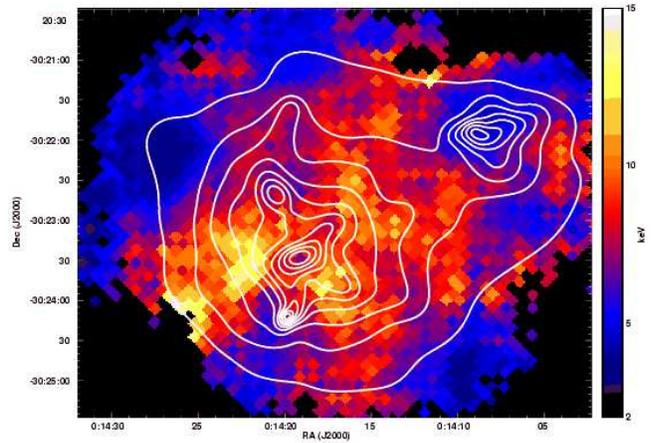}}
\caption{Temperature map of Abell 2744 made with adaptively
binned spectra.  The pixel size in the image is $6\farcs9$.  The binning
length ranges from $29\farcs3$ in high signal-to-noise areas to $63\arcsec$
in the outer parts of the cluster.  Pixels with negative fractional error
in the temperature of $>15$\% (90\% confidence) have been masked out for
clarity.  Contours from Figure~\ref{fig:smoimg} are shown for reference.
\label{fig:tmap}}
\end{figure}

As might be expected from the complex structure in the X-ray image, the
temperature structure in the cluster is quite complex as well.
Figure~\ref{fig:tmap} shows the temperature structure of the cluster in
the higher signal-to-noise areas of the data.  The temperature map was
created using the adaptive binning algorithm described in
\citet{hd00}.  The algorithm bins the data to a minimum of 800 counts
in each extracted spectrum, while also minimizing the sizes of the
extraction regions.  The maximum allowed size of an extraction region
was 63\arcsec$\times$63\arcsec.  Regions of this size with fewer than
800 counts were excluded from this analysis.  The response matrices
were generated on a 32$\times$32 grid in chip coordinates.  The
ancillary responses were generated on a 16$\times$16 grid in the
coordinates of the binned output image.  The systematic errors
introduced by this binning are much smaller than the statistical errors
in the temperature measurements.  The systematic error is approximately
1\%.  Each pixel in the output map is 6\farcs9$\times$6\farcs9.  Each
spectrum was corrected for background using a spectrum from a matching
region in the blank sky background field discussed in \S\ref{sec:obs}.
Regions containing point sources were eliminated from the source and
background event lists.  For clarity, we masked out all pixels in the
final output image with negative fractional errors $> 15$\% (90\%
confidence).

The final image shows several regions of 8--13 keV gas in the core of the
cluster, surrounded by 6--7 keV gas.  The 6--7 keV gas extends out to a
radius of about 740 kpc, and even further in the direction of the northwest
subcluster. Outside that radius, there is some evidence that the
temperature drops even further, although the photon statistics are only
good enough to make such a determination in one small area to the
northeast.  The upper limit on the temperature in this region is nowhere
greater than 5 keV, whereas the lower limit at smaller radii is nowhere
less than about 5.5 keV, so the temperature difference is real.  However,
deeper observations with a larger field of view are needed to confirm this
drop in temperature at large radii, particularly since the temperature drop
to the northeast could be a local phenomenon given the patchiness of the
temperature structure throughout the cluster.

Two fingers of cooler gas extend into the center of the cluster.  The
higher contrast one concides with the south surface brightness
``ridge,'' while the lower contrast finger follows the north ridge.
Both are oriented roughly north-south.  In both cases these fingers of
cool emission terminate with the bright peaks of emission in each
ridge.  The regions of hottest temperature in the core of the cluster
correspond to the other surface brightness ridges.  While ridges
corresponding to the hotter gas, particularly the southwest and
northwest ridges, are relatively sharp narrow features, the cooler
ridges are somewhat broader.  The north ridge is especially broad.  We
note, however, that the hottest region, just east of the cluster
center, is not identified with any surface brightness enhancement.
Another hot region, the finger of hot gas extending due south out of
the cluster center, also is uncorrelated with a surface brightness
enhancement.

The gas in the northwest subcluster, as discussed above, is significantly
cooler than the ambient gas surrounding it.

\section{Galaxy Populations and Distribution}
\label{sec:galaxies}

Abell~2744 is a so-called Butcher-Oemler cluster \citep{bo78a,bo78b,bo84},
with an anomalously large fraction of blue galaxies compared to present-day
clusters.  The cluster has a marked deficiency of elliptical galaxies, a
high percentage of spirals, and a normal percentage of S0 galaxies
\citep{cbs+98}.  Two large elliptical galaxies fall at the center of the
galaxy population, just southeast of the peak of the X-ray emission,
althought they are not significantly more luminous than the next brightest
galaxies.  The spatial distribution of the galaxies, in fact, is roughly
centered on these two galaxies rather than on the peak of the X-ray gas,
although neither of these two galaxies is morphologically identified as a
cD \citep{cbs+98}.

To the northwest, in the direction of the subcluster, \citet{and01}
identifies a clump of galaxies which was associated with the X-ray
subcluster.  The X-ray emission from the subcluster is actually somewhat
farther northwest than the concentration of galaxies.  If the subcluster
were falling straight into the main cluster, one would expected the
collisionless galaxies to precede the collisional ICM as ram pressure from
the interaction with the main cluster slows the descent of the subcluster
into the main cluster potential.  The morphology of the X-ray gas is
largely in agreement with this interpretation, as we will discuss in
greater detail in \S\ref{sec:discussion}.

An extensive catalog of galaxy redshifts in Abell 2744 was compiled by
\citet{cs87} and \citet{cbs+98}.  The distribution in redshift space of the
combined sample from both these sources deviates significantly from a
single Gaussian, as noted by \citet{cbs+98}
for their subsample.  With the addition of the sample from \citet{cs87},
the distribution becomes murkier, but is still clearly non-Gaussian.  While
the sample is relatively small---only 72 galaxies---the distribution is
noticeably bimodal.  \citet{gm01} find one peak at $z = 0.3014$ and the
other peak at $z = 0.3148$.  From this we see that the merger has a quite
high velocity along the line of sight: $\Delta cz = 4000$ km s$^{-1}$.
This unusually large merger velocity strongly suggests that the two
subclusters are at or near their closest approach to one another.  The
extremely large line of sight component of the velocity also indicated that
the merger is occurring largely, though not entirely, along the line of
sight.  As we will show in \S\ref{sec:discussion}, a small transverse
component to the merger is necessary to explain the data.

The further north of the two brightest ellipticals, at ${\rm R.A.} =
00\hd14\md20.6\sd$, ${\rm Dec.} = -30^\circ24\md00\sd$, has a redshift of
$z=0.30000\pm0.00033$ \citep{cs87}, which is at the peak of the bluer
component of the distribution.  The further south of the two bright
ellipticals, at ${\rm R.A.}=00\hd14\md22.0\sd$, ${\rm Dec.} =
-30^\circ24\md20\sd$, has a redshift of $z=0.31870\pm0.00033$ \citep{cs87},
which is near the peak of the redder component of the distribution.
\citet*{gm01} also found that this combined set of galaxies was well fit by
a bimodal distribution, with velocity dispersions of $1121^{+176}_{-88}$ km
s$^{-1}$ and $682^{+97}_{-75}$ km s$^{-1}$, respectively, for the bluer and
redder components.  These velocity dispersions correspond to
$8.0^{+2.7}_{-1.2}$ keV and $3.0^{+0.9}_{-0.7}$ keV.  \citet{gm01} also fit
these redshifts with a single Gaussian with a velocity dispersion that
implies a virial temperature of $\sim$20 keV, more than twice the observed
temperature through most of the cluster.  They also note that the X-ray and
lensing masses for the cluster are highly discrepant, which further
suggests that the cluster is out of equilibrium and that a single cluster
model for the galaxy velocities is therefore inappropriate.

\begin{figure}
\centerline{\epsfig{width=8.5cm,file=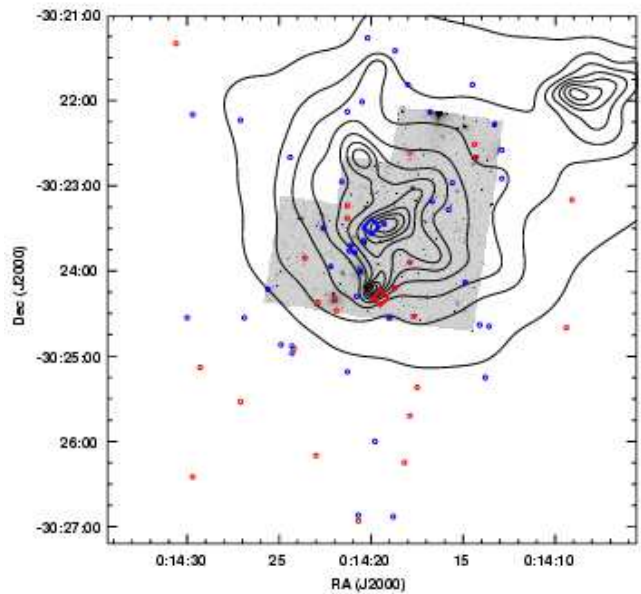}}
\caption{Spatial distribution of cluster members
plotted on and around the HST image of the cluster. Blue circles are
cluster galaxies with $z\leq0.31$. Red circles are cluster galaxies with
$z>0.31$. The blue and red diamonds indicate the centroids of the respective
samples of galaxies. Contours from Figure~\ref{fig:smoimg} are shown for
reference. \label{fig:zhist_img}}
\end{figure}

The two populations in redshift space are also somewhat segregated on
the sky.  Figure~\ref{fig:zhist_img} shows the spatial distribution of
galaxies from the two populations, using $z=0.31$ as the dividing
line.  The set of galaxies with measured redshifts is not uniformly
sampled in all directions from the cluster center.  In order to have a
sample with equal north-south and east-west extent, we took only those
galaxies within 2\farcm7 of the center of the field used by
\citet{cs87}.  This eliminated two galaxies with $z<0.31$ and three
with $z>0.31$ from our sample.  Like the two large ellipticals, the
bluer galaxies tend to be segregated to the north, while the redder
galaxies tend to be segregated to the south.  The mean positions of the
two populations, plotted as diamonds in Figure~\ref{fig:zhist_img},
illustrate the spatial segregation.  A Kolmogorov-Smirnov test confirms
this north-south segregation of the two populations with $>90$\%
confidence.  Thus, while the data are suggestive of a spatial
segregation, they are by no means conclusive.  We should also note that
this sample of galaxies only covers the central 900 kpc, so any cluster
galaxies at larger distances could alter the spatial distribution.  We
suspect that the true centroid of the approaching galaxies, i.e. those
with $z < 0.31$, would in fact change given a more complete sample of
galaxies from a larger field, since the centroid we measure is so close
to the center of the field and since they fill the field so
completely.  The receding galaxies, on the other hand, are a smaller
population on the whole, so we have probably sampled a larger
percentage of it within our field of view.

The only two systematic catalogues of redshifts for this cluster have
concentrated on the central few arcminutes of the cluster and have not
included the northwest subcluster.  Consequently, it is not possible to
determine the line of sight velocity of the northwest clump of galaxies
relative to the main cluster, or even to confirm based on velocity that
they are a distinct population belonging to a separately evolved cluster.
A clue to determining their identity comes instead from their luminosity
function.  The faint end of the luminosity function of the northwest clump
is flatter than that of the main cluster \citep{and01}.  This larger dwarf
fraction is consistent with the galaxies having evolved in a less dense
environment than that of the main cluster \citep{oem74,dre78,lyb+97}.  We
therefore confirm with some certainty the association of the northwest
clump of galaxies with the subcluster detected in X-rays.

\section{Discussion}
\label{sec:discussion}

\subsection{Dynamical History}
\label{ssec:dynamics}

The discussion that follows is an attempt to form a consistent picture for
the dynamical history of the merger based on a confluence of the X-ray,
optical, and radio data.

The X-ray brightness and temperature structure of the main cluster are
extremely complex, particularly in the central 0.5 Mpc.  The ridges in the
X-ray brightness emission are largely correlated with features in the
temperature structure.  The cooler ridges are both oriented north-south,
along the same axis as the centroids of the two galaxy populations
discussed in \S\ref{sec:galaxies}.  Both ridges show some curvature, and in
opposite directions, consistent with being the wakes of the two subclusters
if their interaction has a non-zero impact parameter.  Futhermore, these
two ridges show secondary surface brightness peaks, each about $0\farcm75$
(250 kpc) from the central peak of the main cluster, and each significantly
cooler than the surrounding gas.  These secondary brightness peaks with
extended ridges of emission trailing away from the cluster center combined
with the cold temperature of this gas relative to the rest of the cluster
lead us to interpret the cooler ridges of bright emission to be the wakes
of cooler gas from the cores of the respective subclusters, stripped by ram
pressure through the earlier stages of the merger.  Similarly, we interpret
the two brightness peaks to be the cool cores of the respective
subclusters.  The fact that the cool wakes are visible at all demonstrates
that the merger is not occurring entirely along the line of sight, but must
have at least a small transverse component.

The hotter ridges are seen in the region where the gas should be the most
strongly compressed ahead of the moving cool cores of the two subclusters.
The velocity of the merger derived from the galaxy redshifts implies a Mach
number for the merger of $\sim$2.6, using the temperature of the ambient
cluster gas determined at the radius of the subcluster in
\S\ref{ssec:subcluster} (9.3 keV).  This Mach number is actually a lower
limit since the temperature has probably been boosted by merger shocks by a
factor of $\sim$1.5--2 \citep{rs01} compared to its original value, and
since the velocity derived from the galaxy redshifts excludes the
transverse component of the velocity.  Both of these factors would increase
the actual strength of the shocks produced in the merger.  In any case, the
temperature jump across a $\mathcal{M} = 2.6$ merger shock would be a
factor of 3.  A fit to the northern cool core using a two-temperature MEKAL
model \citep*{kaa92,log95}, with the hotter temperature fixed to that of
the surrounding hot gas, finds a temperature of the cool core of
$4.6^{+3.0}_{-2.7}$ keV compared to a temperature of the surrounding hot
gas of $10.6^{+4.6}_{-2.5}$.  This temperature contrast is consistent with
the predicted shock heating to within the (sizeable) errors.  The two
regions of hot gas which are not correlated with surface brightness
enhancements (see \S\ref{ssec:tmap}) could be due to shocks that are
largely perpendicular to the line of sight, and therefore do not appear as
sharp features in the X-ray image.

The cool core to the north of the cluster center (the south-moving
subcluster) is both larger and brighter than the south core.  This suggests
that it is the more massive of the two.  This assertion may be corroborated
by the galaxy populations: \citet{gm01} found that the bluer population of
galaxies, which has its centroid near the main X-ray peak, has a higher
velocity dispersion than the redder population, which has its centroid
further south.  The mass ratio derived from the velocity dispersions is
$\sim$4:1.  Assuming that the gas has not yet decoupled from the dark
matter and the galaxies, we would expect the positions of the cool cores
and the galaxies to be correlated.  This is probably a safe assumption
since the merger appears to be at an earlier stage than, say, Abell 3667,
in which the gas and dark matter are still coupled \citep{vm02}.  The data
are inconclusive here, however.  The projected separation of the south cool
core from the centroid of the redder galaxies is a mere 60 kpc, but the
separation of the centroid of the bluer galaxies from the north cool core
is 250 kpc.  These centroids are biased, however, by the relatively small
area on the sky over which the galaxies have redshifts available in the
literature.  We therefore conclude that the spatial distributions of the
galaxies are a poor test for determining the connection between the X-ray
cores and the galaxy populations.  The velocity dispersions are a more
reliable test, however, and they appear to indicate that the north cool
core has a negative line of sight velocity relative to the rest frame of
the cluster, while the south core has a positive velocity along the line of
sight in the same frame.

The cool wakes behind the cores, particularly to the north, give more
insight into the dynamical history of the merger.  The northern wake has
significant curvature, indicating that the merger has a non-zero orbital
angular momentum.  Put another way, this shows that the merger is not head
on, but has a non-zero impact parameter.  Unfortunately, it is difficult to
place even a meaningful lower limit on the value of the impact parameter
since the cool wake of the south subcluster is too short to determine the
transverse component of its direction of motion, and the direction implied
by the wake of the north subcluster brings it into the south subcluster
head on.

\subsection{Radio Halo}
\label{ssec:halo}

The radio halo in Abell 2744 is one of the most luminous and most well
studied \citep{gef+01,gfg+01}.  The bulk of the diffuse radio emission is
centered on the main cluster, with a radius of about 3\arcmin.  The cluster
also hosts a radio relic at a projected distance of almost 2 Mpc from the
cluster center.  We will confine our discussion here to the halo, since the
relic is too far from the ACIS focus for our data to contain many source
photons from that region.

\begin{figure}
\centerline{\epsfig{width=8.5cm,file=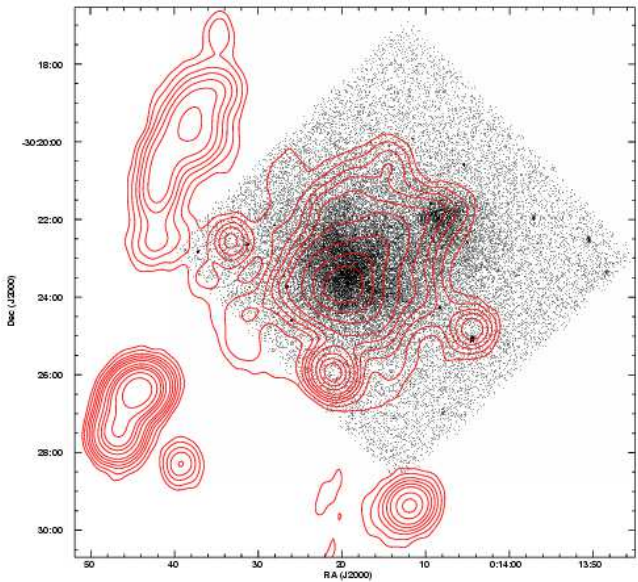}}
\caption{Raw {\it Chandra}\/ image of Abell 2744 with
contours of the radio halo image at 20 cm.  The radio image has been
smoothed with a restoring beam of 50\arcsec$\times$50\arcsec (courtesy of
F.\ Govoni).  Three point sources around the edges of the halo are obvious;
a fourth much fainter point source at ${\rm R.A.} = 00\hd14\md14\sd$,
${\rm Dec} = 30\hd20\md30\sd$ is responsible for the apparent extension of the
halo to the north.
\label{fig:radio}}
\end{figure}

Figure~\ref{fig:radio} shows an image of the radio halo taken with the VLA
at 20 cm, superimposed on the raw {\it Chandra}\/ image.  The offset
between the peaks of the X-ray and radio images noted in \citet{gef+01} is
probably not real, as the peak of the radio emission is resolved into
several smaller peaks at slightly higher resolution than that of our
Figure~\ref{fig:radio} \citep{gfg+01}, the brightest of which is coincident
with the X-ray brightness peak to within a few arcseconds.

The strong correlation between the X-ray and radio surface brightnesses
that was found using lower resolution {\it ROSAT}\/ data by
\citet{gef+01} is also visible in our higher resolution data.  We note
some particularly interesting correlations: all the surface brightness
ridges, both the cool ones and the hot ones, are correlated with
brightness enhancements in the radio.  The northwest subcluster is also
correlated quite strongly with the radio halo.  Also of note is a
strong anti-correlation between the radio brightness ridge to the south
of the cluster and the X-ray brightness.  This feature in the radio
image, however, follows exactly the temperature enhancement in the same
region which we discussed above.  If the high temperature of this gas
is indeed due to a shock in that region, then the enhanced radio
brightness there is most likely the result of current acceleration of
cosmic ray electrons by the shock.  Similarly, the other hot regions,
which are more clearly indicative of shocks from their X-ray
brightnesses, show enhanced radio emission.  Given the large Mach
number inferred above for the merger, the merger shocks should be
strong enough to accelerate electrons to the energies necessary to
produce the observed radio emission \citep{gb03}.

The cool wake from the northern (south-moving) subcluster also shows
enhanced radio emission.  (The south wake shows no enhancement, but the
bright radio point source in that part of the image makes it impossible to
rule out enhanced diffuse emission.)  Again, this probably indicates that
cosmic ray electrons are currently be accelerated in these regions.  No
shocks are likely to exist in the cool stripped gas, so some other
mechanism of particle acceleration is needed.  \citet*{fts03} demonstrated
that turbulent resonant acceleration can generate the necessary electrons
to produce radio halo emission, as long as a population of
trans-relativistic electrons is already present.  Such a population is
clearly present, as indicated by the presence of large-scale diffuse radio
emission.  However, the velocity of the merger in Abell 2744 is so large
that the timescale for the persistence of turbulence is small compared to
the time required to re-accelerate these electrons if the radio emitting
electrons have $\gamma \sim 10^4 - 10^5$.  Electrons with $\gamma \ga
10^3$ have radiative lifetimes of $\la 10^9$ years \citep{sar99a},
while the turbulent timescale is at most $\sim 5 \times 10^8$ yr and
perhaps even $\la 10^8$ yr according to the simulations of
\citet{fts03}, assuming the component subcluster masses derived by
\citet{gm01}.  It is also possible that the enhanced emission in the cool
stripped gas is not due to current particle acceleration, but to a weaker
magnetic field than is present in the rest of the cluster, thereby reducing
the rate of synchrotron losses of the electrons.  This could be the result
of a stretching of the magnetic field as the gas is stripped from the core
of the subcluster.  Unfortunately, the actual mechanism for enhancing the
radio emission cannot be distinguished using the currently available data.
A direct detection of the magnetic field could be made from a spatially
resolved detection of inverse Compton emission, which would enable us to
determine if the enhancement is due to current particle acceleration or to
a locally weaker magnetic field\footnote{This assumes both that variations
in the magnetic field strength are resolved and that the relativistic
electrons are distributed similarly to the field lines.  Coherent
structures over 10s of kiloparsecs observed in Faraday rotation maps
\citep[e.g.][]{eo02} give some hope for the former, but the latter remains
a tenuous assumption.}.  Either of these mechanisms, however, will have the
same observable effect in the radio: the spectral index in the regions of
enhanced emission should be flatter than that in the rest of the cluster
because the electrons will have suffered fewer synchrotron losses.  This
should be easily measurable using lower frequency data with the same
spatial resolution as the 20 cm data.

The northwest subcluster shows further evidence for particle acceleration,
or more likely, re-acceleration.  As can be seen in Figure~\ref{fig:radio},
the radio halo emission extends to the northwest to completely cover the
region of the northwest subcluster.  Two possible scenarios can explain
this extension of the radio halo to the northwewst.

The first scenario is simple shock acceleration from the observed bow shock
ahead of the infalling subcluster.  Other infalling subclusters with Mach
numbers as small as that observed in Abell 2744 do not show any evidence of
diffuse radio emission \citep[e.g. Abell 85,][]{ksr02}.  \citet{gb03}
demonstrated that even in the case of a cluster with a pre-existing
population of suprathermal electrons, weak shocks such as the bow shock in
question produce an energy spectrum of accelerated electrons that is too
steep to explain the observed radio emission.  This predicted lack of radio
emission is consistent with observations of other infalling subclusters,
and thus we find it unlikely that the bow shock in Abell 2744 is
responsible for producing the observed radio emission.

The second scenario assumes that a pool of ``seed'' electrons at mildly
relativistic energies exists, which can be accelerated to the necessary
energies.  A population of suprathermal electrons does indeed exist in much
of the cluster, as the existence of the radio halo demonstrates, but its
presence at radii beyond the edge of the 20 cm radio emission is less
obvious.  As has been seen in other clusters with radio halos, the halos'
spectra steepen with radius, so the halos appear much larger at lower
frequencies \citetext{e.g. Coma, \citealp{gfv+93,drl+97}; for a theoretical
explanation, see \citealp{bsf+01}}.  Therefore it is quite likely that a
population of electrons exists at the radius of the infalling subcluster,
the energies of which are too low to emit synchrotron radiation at 20 cm.
Unfortunately, the lack of data at longer wavelengths for Abell 2744 makes
it impossible to verify this assumption at present.  As long as 2744 is not
unique, it should have the necessary seed electrons at the radius of the
subcluster.  The model of \citet{fts03}, discussed above, is most efficient
at re-accelerating electrons in exactly this sort of situation, i.e.\ a
large mass ratio merger at an early stage where the merger velocity is
still small.  Thus, turbulent re-acceleration of seed electrons could also
account for the extent of the halo across the subcluster.  In principle,
these two scenarios could be distinguished by detailed spectral index maps
of the halo in the vicinity of the subcluster.  In the first scenario, the
halo would show a spectral index gradient from the flattest part near the
bow shock, where the electrons are currently being accelerated, to the wake
of the subcluster, where the electrons would be passively aging.  The
second scenario would create a more uniform spectral index, since electrons
are being accelerated throughout the wake of the subcluster.

\section{Summary}
\label{sec:summary}

We have presented a new {\it Chandra}\/ observation of Abell 2744 which
shows that the main cluster is in a highly disturbed state.  Temperature
and surface brightness variations are observed on all scales along with the
cool cores of the constituent subclusters amid strong ($\mathcal{M} \ga
2$) merger shocks.  The bi-modal distribution of the member galaxies and
the morphology of the radio halo provide further evidence that the cluster
is undergoing a major merger.  We propose a dynamical scenario for the
merger which involves a merger of two subclusters with a mass ratio near
unity and a non-zero impact parameter.  A significant component of the
merger axis is estimated to be along the line of sight.

We have also studied the small merging subcluster to the northwest and
estimate an infall velocity of $\mathcal{M} \sim 1.2$.  We also demonstrate
that this subcluster is not responsible for the bulk of the disturbed
nature of the ICM of the main cluster.  Nonetheless, its effect on the
cluster's extremely powerful radio halo is significant, at least in the
immediate vicinity of the subcluster.  We conclude that turbulent
re-acceleration of electrons in the wake of the subcluster is probably
responsible for the extension of the radio halo across the subcluster.
Future radio observations of the subcluster at lower frequencies should be
able to determined or strongly constrain the formation mechanism of the
radio halo.

\vspace{1cm}
\noindent{\textbf{Acknowledgments}\\
We thank Federica Govoni for the deep radio image of the cluster.  Support
for this work was provided by the National Aeronautics and Space
Administration through {\it Chandra}\/ Award Number GO2-3171X issued by the
{\it Chandra}\/ X-ray Observatory Center, which is operated by the
Smithsonian Astrophysical Observatory for and on behalf of NASA under
contract NAS8-39073.

\bibliographystyle{mn2e}
\bibliography{xray}

\begin{thebibliography}{}

\bibitem[\protect\citeauthoryear{{Abell}}{{Abell}}{1958}]{abe58}
{Abell} G.~O.,  1958, \apjs, 3, 211

\bibitem[\protect\citeauthoryear{{Andreon}}{{Andreon}}{2001}]{and01}
{Andreon} S.,  2001, \apj, 547, 623

\bibitem[\protect\citeauthoryear{{Bautz} \& {Morgan}}{{Bautz} \&
  {Morgan}}{1970}]{bm70}
{Bautz} L.~P.,  {Morgan} W.~W.,  1970, \apjl, 162, L149

\bibitem[\protect\citeauthoryear{{Brunetti}, {Setti}, {Feretti} \&
  {Giovannini}}{{Brunetti} et~al.}{2001}]{bsf+01}
{Brunetti} G.,  {Setti} G.,  {Feretti} L.,    {Giovannini} G.,  2001, \mnras,
  320, 365

\bibitem[\protect\citeauthoryear{{Butcher} \& {Oemler}}{{Butcher} \&
  {Oemler}}{1978a}]{bo78a}
{Butcher} H.,  {Oemler} A.,  1978a, \apj, 219, 18

\bibitem[\protect\citeauthoryear{{Butcher} \& {Oemler}}{{Butcher} \&
  {Oemler}}{1978b}]{bo78b}
{Butcher} H.,  {Oemler} A.,  1978b, \apj, 226, 559

\bibitem[\protect\citeauthoryear{{Butcher} \& {Oemler}}{{Butcher} \&
  {Oemler}}{1984}]{bo84}
{Butcher} H.,  {Oemler} A.,  1984, \apj, 285, 426

\bibitem[\protect\citeauthoryear{{Couch}, {Barger}, {Smail}, {Ellis} \&
  {Sharples}}{{Couch} et~al.}{1998}]{cbs+98}
{Couch} W.~J.,  {Barger} A.~J.,  {Smail} I.,  {Ellis} R.~S.,    {Sharples}
  R.~M.,  1998, \apj, 497, 188

\bibitem[\protect\citeauthoryear{{Couch} \& {Newell}}{{Couch} \&
  {Newell}}{1984}]{cn84}
{Couch} W.~J.,  {Newell} E.~B.,  1984, \apjs, 56, 143

\bibitem[\protect\citeauthoryear{{Couch} \& {Sharples}}{{Couch} \&
  {Sharples}}{1987}]{cs87}
{Couch} W.~J.,  {Sharples} R.~M.,  1987, \mnras, 229, 423

\bibitem[\protect\citeauthoryear{{Deiss}, {Reich}, {Lesch} \&
  {Wielebinski}}{{Deiss} et~al.}{1997}]{drl+97}
{Deiss} B.~M.,  {Reich} W.,  {Lesch} H.,    {Wielebinski} R.,  1997, \aap, 321,
  55

\bibitem[\protect\citeauthoryear{{Dressler}}{{Dressler}}{1978}]{dre78}
{Dressler} A.,  1978, \apj, 223, 765

\bibitem[\protect\citeauthoryear{{Ebeling}, {Voges}, {Bohringer}, {Edge},
  {Huchra} \& {Briel}}{{Ebeling} et~al.}{1996}]{evb+96}
{Ebeling} H.,  {Voges} W.,  {Bohringer} H.,  {Edge} A.~C.,  {Huchra} J.~P.,
  {Briel} U.~G.,  1996, \mnras, 281, 799

\bibitem[\protect\citeauthoryear{{Eilek} \& {Owen}}{{Eilek} \&
  {Owen}}{2002}]{eo02}
{Eilek} J.~A.,  {Owen} F.~N.,  2002, \apj, 567, 202

\bibitem[\protect\citeauthoryear{{Ettori} \& {Fabian}}{{Ettori} \&
  {Fabian}}{2000}]{ef00b}
{Ettori} S.,  {Fabian} A.~C.,  2000, \mnras, 317, L57

\bibitem[\protect\citeauthoryear{{Fujita}, {Takizawa} \& {Sarazin}}{{Fujita}
  et~al.}{2003}]{fts03}
{Fujita} Y.,  {Takizawa} M.,    {Sarazin} C.~L.,  2003, \apj, {584}, 190

\bibitem[\protect\citeauthoryear{{Gabici} \& {Blasi}}{{Gabici} \&
  {Blasi}}{2003}]{gb03}
{Gabici} S.,  {Blasi} P.,  2003, \apj, 583, 695

\bibitem[\protect\citeauthoryear{{Giovannini}, {Feretti}, {Venturi}, {Kim} \&
  {Kronberg}}{{Giovannini} et~al.}{1993}]{gfv+93}
{Giovannini} G.,  {Feretti} L.,  {Venturi} T.,  {Kim} K.-T.,    {Kronberg}
  P.~P.,  1993, \apj, 406, 399

\bibitem[\protect\citeauthoryear{{Giovannini}, {Tordi} \&
  {Feretti}}{{Giovannini} et~al.}{1999}]{gtf99}
{Giovannini} G.,  {Tordi} M.,    {Feretti} L.,  1999, New Astronomy, 4, 141

\bibitem[\protect\citeauthoryear{{Girardi} \& {Mezzetti}}{{Girardi} \&
  {Mezzetti}}{2001}]{gm01}
{Girardi} M.,  {Mezzetti} M.,  2001, \apj, 548, 79

\bibitem[\protect\citeauthoryear{{Govoni}, {En{\ss}lin}, {Feretti} \&
  {Giovannini}}{{Govoni} et~al.}{2001}]{gef+01}
{Govoni} F.,  {En{\ss}lin} T.~A.,  {Feretti} L.,    {Giovannini} G.,  2001,
  \aap, 369, 441

\bibitem[\protect\citeauthoryear{{Govoni}, {Feretti}, {Giovannini}, {B{\"
  o}hringer}, {Reiprich} \& {Murgia}}{{Govoni} et~al.}{2001}]{gfg+01}
{Govoni} F.,  {Feretti} L.,  {Giovannini} G.,  {B{\" o}hringer} H.,  {Reiprich}
  T.~H.,    {Murgia} M.,  2001, \aap, 376, 803

\bibitem[\protect\citeauthoryear{{Houck} \& {Denicola}}{{Houck} \&
  {Denicola}}{2000}]{hd00}
{Houck} J.~C.,  {Denicola} L.~A.,  2000, in ASP Conf. Ser. 216: Astronomical
  Data Analysis Software and Systems IX {ISIS: An Interactive Spectral
  Interpretation System for High Resolution X-Ray Spectroscopy}.
{ASP}, {San Francisco}, p.~591

\bibitem[\protect\citeauthoryear{{Kaastra}}{{Kaastra}}{1992}]{kaa92}
{Kaastra} J.~S.,  1992, Technical report, {An X-Ray Spectral Code for Optically
  Thin Plasmas}

\bibitem[\protect\citeauthoryear{{Kempner}, {Sarazin} \& {Ricker}}{{Kempner}
  et~al.}{2002}]{ksr02}
{Kempner} J.~C.,  {Sarazin} C.~L.,    {Ricker} P.~M.,  2002, \apj, 579, 236

\bibitem[\protect\citeauthoryear{{Liedahl}, {Osterheld} \&
  {Goldstein}}{{Liedahl} et~al.}{1995}]{log95}
{Liedahl} D.~A.,  {Osterheld} A.~L.,    {Goldstein} W.~H.,  1995, \apjl, 438,
  115

\bibitem[\protect\citeauthoryear{{Lopez-Cruz}, {Yee}, {Brown}, {Jones} \&
  {Forman}}{{Lopez-Cruz} et~al.}{1997}]{lyb+97}
{Lopez-Cruz} O.,  {Yee} H.~K.~C.,  {Brown} J.~P.,  {Jones} C.,    {Forman} W.,
  1997, \apjl, 475, L97

\bibitem[\protect\citeauthoryear{{Markevitch}, {Gonzalez}, {David},
  {Vikhlinin}, {Murray}, {Forman}, {Jones} \& {Tucker}}{{Markevitch}
  et~al.}{2002}]{mgd+02}
{Markevitch} M.,  {Gonzalez} A.~H.,  {David} L.,  {Vikhlinin} A.,  {Murray} S.,
   {Forman} W.,  {Jones} C.,    {Tucker} W.,  2002, \apjl, 567, L27

\bibitem[\protect\citeauthoryear{{Markevitch} et~al.}{{Markevitch}
  et~al.}{2000}]{mpn+00}
{Markevitch} M.,  {Ponman} T.~J.,  {Nulsen} P.~E.~J.,  {Bautz} M.~W.,  {Burke}
  D.~J.,  {David} L.~P.,  {Davis} D.,  {Donnelly} R.~H.,  {Forman} W.~R.,
  {Jones} C.,  {Kaastra} J.,  {Kellogg} E.,  {Kim} D.-W.,  {Kolodziejczak} J.,
  {Mazzotta} P.,  {Pagliaro} A.,  {Patel} S.,  {Van Speybroeck} L.,
  {Vikhlinin} A.,  {Vrtilek} J.,  {Wise} M.,    {Zhao} P.,  2000, \apj, 541,
  542

\bibitem[\protect\citeauthoryear{{Markevitch} \& {Vikhlinin}}{{Markevitch} \&
  {Vikhlinin}}{2001}]{mv01}
{Markevitch} M.,  {Vikhlinin} A.,  2001, \apj, 563, 95

\bibitem[\protect\citeauthoryear{{Moekel}}{{Moekel}}{1949}]{moe49}
{Moekel} W.~E., , 1949, {Approximate Method for Predicting Forms and Location
  of Detached Shock Waves Ahead of Plane or Axially Symmetric Bodies}, NACA
  Technical Note 1921

\bibitem[\protect\citeauthoryear{{Oemler}}{{Oemler}}{1974}]{oem74}
{Oemler} A.~J.,  1974, \apj, 194, 1

\bibitem[\protect\citeauthoryear{{Ricker} \& {Sarazin}}{{Ricker} \&
  {Sarazin}}{2001}]{rs01}
{Ricker} P.~M.,  {Sarazin} C.~L.,  2001, \apj, 561, 621

\bibitem[\protect\citeauthoryear{{Sarazin}}{{Sarazin}}{1999}]{sar99a}
{Sarazin} C.~L.,  1999, \apj, 520, 529

\bibitem[\protect\citeauthoryear{{Vikhlinin}, {Markevitch} \&
  {Murray}}{{Vikhlinin} et~al.}{2001a}]{vmm01b}
{Vikhlinin} A.,  {Markevitch} M.,    {Murray} S.~S.,  2001a, \apj, 551, 160

\bibitem[\protect\citeauthoryear{{Vikhlinin}, {Markevitch} \&
  {Murray}}{{Vikhlinin} et~al.}{2001b}]{vmm01a}
{Vikhlinin} A.,  {Markevitch} M.,    {Murray} S.~S.,  2001b, \apjl, 549, L47

\bibitem[\protect\citeauthoryear{{Vikhlinin} \& {Markevitch}}{{Vikhlinin} \&
  {Markevitch}}{2002}]{vm02}
{Vikhlinin} A.~A.,  {Markevitch} M.~L.,  2002, Astronomy Letters, 28, 508

\end{thebibliography}

\end{document}